\documentclass[acmsmall,screen,nonacm,review=false,timestamp=false]{acmart}

\AtBeginDocument{%
  }

\setcopyright{acmlicensed}
\copyrightyear{2025}
\acmYear{2025}
\acmDOI{XXXXXXX.XXXXXXX}

\acmConference[Conference acronym 'XX]{Make sure to enter the correct
  conference title from your rights confirmation emai}{June 03--05,
  2018}{Woodstock, NY}

\usepackage{booktabs} 
\usepackage{multirow}
\usepackage{colortbl}
\usepackage{enumitem}
\usepackage{xcolor}

\definecolor{verylightgray}{rgb}{0.92,0.92,0.92}
\newcommand{\bc}[1]{\cellcolor{verylightgray}#1}

\newcommand{\dg}{$^\dagger$} 
\newcommand{\sd}{$^*$} 
\newcommand{\lw}{$^\triangledown$} 

\usepackage{amsmath}
\usepackage[capitalise]{cleveref}

\newcommand{\hide}[1]{\iffalse {#1} \fi}

\everypar\expandafter{\the\everypar\looseness -1}
\begin{document}

\title{Exploring the Effectiveness of Multi-stage Fine-tuning for Cross-encoder Re-rankers}

\author{Francesca Pezzuti}
\orcid{0009-0005-2364-2043}
\email{francesca.pezzuti@phd.unipi.it}
\affiliation{
\institution{University of Pisa}
  \city{Pisa}
  \country{Italy}
}
\author{Sean MacAvaney}
\orcid{0000-0002-8914-2659}
\email{sean.macavaney@glasgow.ac.uk}
\affiliation{
\institution{University of Glasgow}
  \city{Glasgow}
  \country{UK}
}
\author{Nicola Tonellotto}
\orcid{0000-0002-7427-1001}
\email{nicola.tonellotto@unipi.it}
\affiliation{
\institution{University of Pisa}
  \city{Pisa}
  \country{Italy}
}          

\begin{abstract}
State-of-the-art cross-encoders can be fine-tuned to be highly effective in passage re-ranking. 
The typical fine-tuning process of cross-encoders as re-rankers requires large amounts of manually labelled data, a \textit{contrastive learning} objective, and a set of heuristically sampled negatives. An alternative recent approach for fine-tuning instead involves teaching the model to mimic the rankings of a highly effective large language model using a \textit{distillation} objective. These fine-tuning strategies can be applied either individually, or in sequence.
In this work, we systematically investigate the effectiveness of point-wise cross-encoders when fine-tuned independently in a single stage, or sequentially in two stages. 
Our experiments show that the effectiveness of point-wise cross-encoders fine-tuned using contrastive learning is indeed on par with that of models fine-tuned with multi-stage approaches. Code is available for reproduction at \url{https://github.com/fpezzuti/multistage-finetuning}.

\end{abstract}

\maketitle   

\section{Introduction}
    With the introduction of contextualised language models such as BERT~\cite{devlin2019bert}, ELECTRA~\cite{clark2020electra}, and RoBERTa~\cite{liu2019roberta}, a new family of highly effective neural Information Retrieval (IR) systems quickly emerged. Within the wide range of \textit{neural IR models}, which includes bi-encoders~\cite{izacard2022contriever,xiong2021ance,hofstatter2021tasb} and late-interaction models~\cite{khattab2020colbert}, a significant category is formed by cross-encoders like monoBERT~\cite{nogueira2019monoBERT} and monoT5~\cite{nogueira2020monoT5}.
    These cross-encoders leverage pre-trained language models to estimate the relevance between a query and a document by jointly encoding them in a shared latent representation that effectively captures semantic interactions.

    However, before being used as rankers, pre-trained cross-encoders must be fine-tuned for the task.
    Over the years, various fine-tuning techniques have been proposed to this purpose. 
    The \textit{vanilla} method ~\cite{Nogueira2019PassageRW} adopts a Binary Cross-Entropy (BCE) loss to frame the problem of estimating query-document relevance as a binary classification task.
    Despite being effective at predicting relevance, by making independent relevance predictions, models fine-tuned with BCE have a binary understanding of relevance and fail at estimating relative rankings.
    In contrast, \textit{contrastive learning} techniques address these shortcomings by relying on heuristically selected negatives, i.e., non-relevant documents, allowing the re-ranker to learn to assign higher scores to relevant documents compared to non-relevant ones. One widely used contrastive learning loss is the Noise Contrastive Estimation (NCE) loss~\cite{gutmann2010NCE}, which takes into account randomly selected negatives.
    To enhance robustness and effectiveness, contrastive learning losses often incorporate \textit{hard negatives}, i.e., non-relevant passages closely related to the query. The Localized Contrastive-Estimation (LCE) loss~\cite{gao2021LCE} is an effective variant of NCE that uses hard negatives randomly sampled from the ranking lists of a retriever.
    
    While cross-encoders exhibit remarkable effectiveness when fine-tuned as rankers, they are computationally expensive. Consequently, they are often used as re-rankers in retrieve-then-rerank systems to refine the ranking of a small subset of documents initially induced by a more efficient model like BM25~\cite{robertson1994bm25}, which serves as retriever.
    
    With advancements in large language models, a new list-wise re-ranking paradigm has emerged, such as RankGPT~\cite{sun2023RankGPT} and LEAF~\cite{bevilacqua2022autoregressive}. These models use large language models (LLMs) to re-rank a given set of documents with respect to one another. Although these models excel at ranking, they are more computationally intensive than cross-encoders, and often incur significant monetary costs due to the use of proprietary LLMs~\cite{rraml}.

    However, both computational demands and monetary costs can be substantially reduced using the \textit{knowledge distillation} paradigm~\cite{Hinton2015DistillingTK}. This approach allows simpler models, like cross-encoders, to capture the capabilities of complex models, like generative rankers.
    In the context of IR, distillation involves fine-tuning a smaller ranker, known as \textit{student}, to mimic the rankings produced by a highly effective but expensive model, referred to as \textit{teacher}. During this process, the student learns from the soft labels derived from the ranking list generated by the teacher.   
    One widely used ranking distillation loss is RankNet~\cite{burges2005ranknet}, which aims to minimise the number of incorrect relative document orders between the ranking generated by student and teacher.
   
    While contrastive learning and distillation are typically applied separately, Shlatt et al. applied  to cross-encoders contrastive learning with LCE loss, followed by distillation with RankNet~\cite{schlatt2024setencoder} or the Approx. Discounted Rank MSE loss~\cite{schlatt2024systematicinvestigation}. However, they only explored this particular sequence, and found no significant effectiveness improvements over single-stage fine-tuning.
    To the best of our knowledge, Schlatt et al. are the only ones who applied a multi-stage fine-tuning strategy to cross-encoder re-rankers. Yet, the cumulative impact on effectiveness of sequentially applying contrastive learning and distillation to cross-encoders has not been fully explored.
    
    In this work, we aim to fill this gap by systematically evaluating the effectiveness of cross-encoders fine-tuned with either a single-stage approach -- contrastive learning or distillation -- or a multi-stage approach, combining both. Our findings reveal that there is no significant improvement in effectiveness when fine-tuning cross-encoders with a multi-stage approach compared to using single-stage fine-tuning.
\section{Background \& Methodology}

    Let $q$ denote a textual query, and $\mathcal{D}$ a corpus of textual documents. Let $\mathcal{R}_q^k=\left \{d_1, \ldots, d_k \right \}$ with $d_i \in \mathcal{D}$, denote the set of top $k$ documents retrieved by the retriever for $q$. In a multi-stage retrieval system, given $q$ and $\mathcal{R}_q^k$, the re-ranker assigns to each $d_i \in \mathcal{R}_q^k$ a relevance score $s(q,d_i)$ w.r.t. $q$. The relevance scores over $\mathcal{R}_q^k$ are then used to infer a re-ranking of the $k$ candidates.

    To compute these relevance scores, a cross-encoder (CE) leverages a transformer encoder with cross-attention that allows it to capture the interactions between query tokens and document tokens.
    
    However, before pre-trained CEs can be effectively used as re-rankers, they must undergo fine-tuning with \textit{contrastive learning}, \textit{knowledge distillation}, or a combination of both, to optimize the parameters of the CE.

    To apply contrastive learning, each training instance should be formed by a query, a relevant document, and a set of $h$ hard negatives randomly sampled from the ranking list generated by a first-stage ranker.
    Formally, given a query $q$, let $d^+$ denote the relevant document w.r.t. $q$, and let $\mathcal{H}= \left \{ d_1, \ldots d_h \mid d_i \sim \mathcal{R}_q^k \right\}$ be the set of $h$ hard negatives sampled from the training ranking list $\mathcal{R}_q^k$ associated to $q$.
    For the query $q$, the Localized Contrastive-Estimation Loss (LCE) is computed as:    
    \begin{equation*}
        \mathcal{L}_{LCE}(q) = - \log \frac{e^{s(q,d^+)}}{e^{s(q,d^+)} + \sum_{d_i \in \mathcal{H}} e^{s(q, d_i)}} 
    \end{equation*}
    The main limitation of this loss is that it relies on hard labels, meaning that a document is considered either strictly relevant, or non-relevant. In particular, LCE does not use any rank information from $\mathcal{R}_q^k$, which could serve as soft labels. However, knowledge distillation can address this limitation.
    
    Indeed, when distilling, the ranks $r_i \in [1,\ldots, k]$, assigned by the teacher ranker to $d_i \in \mathcal{R}_q^k$ when generating $\mathcal{R}_q^k$ for a query $q$, are utilised as soft, fine-grained labels. 
    Given this notation, the RankNet loss for a query $q$ is computed as:
    \begin{equation*}
        \mathcal{L}_{RankNet}(q) = \sum_{r_i < r_j} \log \left (1+ e^{s(q,d_i) -s(q,d_j)} \right )
    \end{equation*} with $d_i,d_j \in \mathcal{R}_q^k$.
    However, the effectiveness of the student ranker is closely tied to the quality of the teacher, as its training heavily relies on teacher ranks.

    While contrastive learning and distillation are typically applied separately to CEs, combining them in sequence could potentially create a synergistic effect that enhances performance. Therefore, we aim to explore whether combinations of these techniques can improve the re-ranking effectiveness of CEs.
    
    In the following, we focus on comparing the two different fine-tuning approaches for CE re-rankers, namely using contrastive learning with LCE or knowledge distillation with RankNet. 
    
    Next, we investigate the re-ranking effectiveness of CEs fine-tuned with a combination of these single-stage fine-tuning strategies applied in sequence, to determine which is the best multi-stage approach. Finally, we investigate whether combining the two single-stage strategies sequentially improves effectiveness over using a single-stage fine-tuning.
\section{Experimental Setup}

    We conduct experiments to answer the following research questions:

    \begin{description}[noitemsep, nolistsep]
        \item[RQ1] Which of the presented \textit{single-stage} fine-tuning strategies produces more effective cross-encoder re-rankers?
        \item[RQ2] Which of the presented \textit{multi-stage} fine-tuning strategies produces more effective cross-encoder re-rankers?
        \item[RQ3] Is the best multi-stage strategy from RQ2 more effective than the best single-stage strategy from RQ1?
    \end{description}
    
            In our experiments, we use BM25 and ColBERTv2~\cite{santhanam2022colbertv2} as rankers, using Pyserini to  generate ranking lists. As CE re-rankers, we use  ELECTRA 
            (denoted El. in the following) and RoBERTa (denoted Ro. in the following). 
            We evaluate re-ranking effectiveness using the MS MARCO~\cite{bajaj2016msmarco} collection of $8.8$ million passages and four query sets: DEV SMALL~\cite{bajaj2016msmarco}, TREC DL 19, 20, HARD~\cite{craswell2020dl2019,craswell2021dl2020,mackie2021dlhard}, all loaded via ir-datasets~\cite{macavaney2021irds}.
            We measure AP, nDCG@10, and MRR@10 using ir-measures~\cite{MacAvaney2022irmeasures}, but we omit the cutoff value @10 in the tables.
            For significance testing, we use a two-tailed paired Student's t-test with $p = 0.01$. 
            
            For contrastive learning with LCE (denoted C in the following), we use the dataset\footnote{\href{https://zenodo.org/records/10952882}{https://zenodo.org/records/10952882}} from Schlatt et al.~\cite{schlatt2024dataset}, consisting of the top $500$ passages retrieved by ColBERTv2 for $503k$ MS MARCO train queries. Following prior research~\cite{gao2021LCE,pradeep2022LCEstudy}, during fine-tuning, we randomly sample hard negatives from the top $200$. However, while Gao et al.~\cite{gao2021LCE} use $h=7$, Pradeep et al. \cite{pradeep2022LCEstudy} use up to $h=31$ and observe that increasing $h$ improves effectiveness with no plateauing up to $31$. Hence, we use $h=99$.
            To distill with RankNet (denoted D in the following), we use the dataset\footnote{\href{https://github.com/sunnweiwei/RankGPT}{https://github.com/sunnweiwei/RankGPT}} from Sun et al.~\cite{sun2023RankGPT}, which comprises the top $20$ passages retrieved by RankGPT-3.5 for $100k$ MS MARCO train queries.\footnote{Actually $90.7k$ after our pre-processing and cleaning.} 
            For both C and D, we split the dataset into train ($99\%$) and validation ($1\%$), and use the AdamW optimizer~\cite{Loshchilov2019AdamW}.
            For C, we use a learning rate $lr=10^{-5}$ and we stop after $25k$ steps if applied as first stage, else after $31k$ steps. For D, we use $lr=10^{-5}$ in first stage, stopping after $2k$ steps for El., and $1k$ for Ro; when using D as second-stage, for El. we use $lr=10^{-8}$, stopping after $1k$ steps, for Ro. we use $lr=10^{-9}$ and stop after $3k$ steps.
\section{Results}

    \begin{table}[t!]
\centering
\caption{Re-ranking effectiveness of CEs fine-tuned with contrastive learning (C) or distillation (D). Significant differences between the two fine-tuned versions of the same CE are denoted with \sd\ , statistically significant difference w.r.t. the baseline are denoted with \dg. Bold values denote the best value between two versions of the same CE, while \lw\ denotes values below the baseline. 
} 
\label{tab:onestage}
\resizebox{\columnwidth}{!}{%
\begin{tabular}{cccccccccccccc}

\toprule
\multicolumn{2}{c}{\multirow{2}{*}{Re-rank}} & \multicolumn{3}{c}{DL 19}                                   & \multicolumn{3}{c}{DL 20}                           & \multicolumn{3}{c}{DL HARD}                            & \multicolumn{3}{c}{DEV SMALL}                     \\ 
\cmidrule(lr){3-5}\cmidrule(lr){6-8}\cmidrule(lr){9-11}\cmidrule(lr){12-14}
                     &                         & AP & nDCG        & MRR            & AP                & nDCG        & MRR         &   AP & nDCG        & MRR            & AP                & nDCG        & MRR        \\ 
\midrule
\multicolumn{14}{c}{BM25} \\
\midrule

\multicolumn{2}{c}{\bc --}                       & \bc.3035             & \bc.5121             & \bc.7138       & \bc.2811             & \bc.4769             & \bc.6653             & \bc.4019             & \bc.6744             & \bc.8140             & \bc.4482             & \bc.6716             & \bc.7997              \\
\multirow{2}{*}{El.} & C                    & \textbf{.3651\sd}    & \textbf{.7236\sd\dg} & .8314          & \textbf{.4012\sd\dg} & \textbf{.6759\sd\dg} & \textbf{.8278\dg}    & \textbf{.4461}       & \textbf{.7376\dg}    & \textbf{.8682}       & \textbf{.4984\sd\dg} & \textbf{.7391\sd\dg} & \textbf{.8536\sd\dg}  \\
                     & D                    & .3345                & .6691\dg             & \textbf{.8876} & .3531\dg             & .6147\dg             & .7720                & .3840\lw             & .6775\dg             & .8651                & .4132\lw\dg          & .6417\lw\dg          & .7752\lw\dg           \\
\midrule
\multirow{2}{*}{Ro.} & C                    & \textbf{.3687\sd}    & \textbf{.7356\sd\dg} & \textbf{.8651} & \textbf{.3997\sd\dg} & \textbf{.6720\sd\dg} & \textbf{.8438\sd\dg} & \textbf{.4367}       & \textbf{.7221\dg}    & \textbf{.8411}       & \textbf{.4917\sd\dg} & \textbf{.7383\sd\dg} & \textbf{.8633\sd\dg}  \\
                     & D                    & .3284                & .6558\dg             & .8353          & .3505\dg             & .6029\dg             & .7352                & .1972\lw             & .3652\lw\dg          & .5645\lw             & .2744\lw\dg          & .3242\lw\dg          & .2705\lw\dg           \\
\midrule
\multicolumn{14}{c}{ColBERTv2} \\
\midrule

\multicolumn{2}{c}{\bc --}                       & \bc\textbf{.}5077    & \bc.7369             & \bc.8876       & \bc.5160             & \bc.7328             & 
\bc.8282             & \bc.2641             & \bc.4021             & \bc.5531             & \bc.3956             & \bc.4569             & \bc.3907              \\

\multirow{2}{*}{El.} & C                    & \textbf{.4701\lw\sd} & \textbf{.7537\sd}    & .8663\lw       & \textbf{.5205\sd}    & \textbf{.7337\sd}    & \textbf{.8536\sd}    & \textbf{.2541\lw}    & \textbf{.4022}       & .5150\lw             & \textbf{.4228\sd\dg} & \textbf{.4844\sd\dg} & \textbf{.4191\sd\dg}  \\
                     & D                    & .4072\lw             & .6916\lw\dg          & \textbf{.8915} & .4273\lw\dg          & .6407\lw\dg          & .7780\lw             & .2316\lw             & .3689\lw             & \textbf{.5684}       & .2896\lw\dg          & .3421\lw\dg          & .2794\lw\dg           \\
\midrule
\multirow{2}{*}{Ro.} & C                    & \textbf{.4633\lw\sd} & \textbf{.7333\lw}    & .8391\lw       & \textbf{.5136\lw\sd} & \textbf{.7370\sd}    & \textbf{.8617\sd}    & \textbf{.2638\lw}    & \textbf{.4211}       & \textbf{.5640}       & \textbf{.4151\sd\dg} & \textbf{.4773\sd\dg} & \textbf{.4105\sd\dg}  \\
                     & D                    & .4105\lw\dg          & .6784\lw             & \textbf{.8729} & .4238\lw\dg          & .6369\lw\dg          & .7282\lw             & .2194\lw             & .3660\lw             & .5534                & .2933\lw\dg          & .3441\lw\dg          & .2829\lw\dg       \\
 \bottomrule
\end{tabular} 
} 
\end{table}

\begin{table}[t!]
    \centering
   \caption{Re-ranking effectiveness of CEs fine-tuned with contrastive learning followed by distillation (C$\to$D), or the reverse (D$\to$C). Significant differences between the two fine-tuned versions of the same CE are denoted with \sd, and statistically significant differences w.r.t. the baseline are denoted with \dg. Bold values denote the best value between two versions of the same CE, while \lw\ denotes values below the baseline.
   } 
   \label{tab:twostages}
\resizebox{\columnwidth}{!}{%
\begin{tabular}{@{}clccccccccccccccc@{}}
\toprule
\multicolumn{2}{c}{\multirow{2}{*}{Re-rank}} & \multicolumn{3}{c}{DL 19}                                   & \multicolumn{3}{c}{DL 20}                           & \multicolumn{3}{c}{DL HARD}                            & \multicolumn{3}{c}{DEV SMALL}                     \\ 
\cmidrule(lr){3-5}\cmidrule(lr){6-8}\cmidrule(lr){9-11}\cmidrule(lr){12-14}
                     &                         & AP & nDCG        & MRR            & AP                & nDCG        & MRR         &   AP & nDCG        & MRR            & AP                & nDCG        & MRR        \\ 
\midrule
\multicolumn{14}{c}{BM25} \\
\midrule
\multicolumn{2}{c}{\bc --}     & \bc.3035          & \bc.5121          & \bc.7138          & \bc.2811          & \bc.4769          & \bc.6653          & \bc.1622          & \bc.2886          & \bc.4740          & \bc.1941          & \bc.2301         & \bc.1855          \\
\multirow{2}{*}{El.} & C$\to$D & \textbf{.3652}    & .7234\dg          & \textbf{.8391}    & .4003\dg          & .6775\dg          & .8380\dg          & \textbf{.2085}    & \textbf{.3858}\dg & \textbf{.5184}    & \textbf{.3696}\dg & \textbf{.4209}\dg & .3708\dg          \\
                     & D$\to$C & .3633             & \textbf{.7304}\dg & .8262             & \textbf{.4065}\dg & \textbf{.6822}\dg & \textbf{.8438}\dg & .2050             & .3841\dg          & .5041             & .3693\dg          & .4208\dg          & \textbf{.3712}\dg \\
 \midrule
\multirow{2}{*}{Ro.} & C$\to$D & \textbf{.3685}    & \textbf{.7348}\dg & \textbf{.8651}    & \textbf{.4012}\dg & \textbf{.6752}\dg & .8438\dg          & \textbf{.2228}    & \textbf{.4051}\dg & \textbf{.5657}    & .3667\dg          & .4182\dg          & .3679\dg          \\
                     & D$\to$C & .3628             & .7323\dg          & .8529             & .3998\dg          & .6687\dg          & \textbf{.8525}\dg & .2160             & .3912\dg          & .5338             & \textbf{.3678}\dg & \textbf{.4187}\dg & \textbf{.3693}\dg \\
\midrule
\multicolumn{14}{c}{ColBERTv2} \\
\midrule
\multicolumn{2}{c}{\bc --}     & \bc\textbf{.5077} & \bc.7369          & \bc.8876          & \bc.5160          & \bc.7328          & \bc.8282          & \bc.2641          & \bc.4021          & \bc.5531          & \bc.3956          & \bc.4569          & \bc.3907          \\
\multirow{2}{*}{El.} & C$\to$D & .4705\lw          & .7550             & \textbf{.8740}\lw & .5198             & .7334             & .8638             & .2525\lw          & .4015\lw          & .5137\lw          & .4227\dg          & .4841\dg          & .4182\dg          \\
                     & D$\to$C & \textbf{.4732}\lw & \textbf{.7632}    & .8599\lw          & \textbf{.5265}    & \textbf{.7585}    & \textbf{.8824}    & \textbf{.2552}\lw & \textbf{.4104}    & \textbf{.5254}\lw & \textbf{.4234}\dg & \textbf{.4855}\dg & \textbf{.4193}\dg \\

 \midrule
\multirow{2}{*}{Ro.} & C$\to$D & .4633\lw          & \textbf{.7341}\lw & .8411\lw          & \textbf{.5150}\lw & \textbf{.7375}    & .8617             & \textbf{.2637}\lw & \textbf{.4217}    & \textbf{.5647}    & .4148\dg          & .4771\dg          & .4106\dg          \\
                     & D$\to$C & \textbf{.4646}\lw & .7337\lw          & \textbf{.8510}\lw & .5115\lw          & .7322\lw          & \textbf{.8675}    & .2569\lw          & .4125             & .5278\lw          & \textbf{.4171}\dg & \textbf{.4788}\dg & \textbf{.4132}\dg \\
\bottomrule
\end{tabular}%
} 
\end{table}
    We first explore RQ1: whether is it more effective a cross-encoder fine-tuned with contrastive learning (C) or distillation (D).
    ~\cref{tab:onestage} compares the re-ranking effectiveness of El. and Ro. fine-tuned with C or D.
    Consistently on all query sets, we observe that C generates more effective CEs than D, both at re-ranking BM25 and ColBERTv2 results. Except for DL HARD, differences between C and D are generally statistically significant.
    We also observe that for most benchmarks and metrics, CEs fine-tuned with C are statistically more effective than the baseline, while those fine-tuned with D are often statistically less effective.
    To conclude on RQ1, our experiments show that fine-tuning CEs with contrastive learning is more effective than with knowledge distillation. 
    
    Next, we explore RQ2: whether is it more effective a CE fine-tuned with C followed by D (C$\to$D), or the reverse (D$\to$C).
    \cref{tab:twostages} shows the effectiveness of CEs fine-tuned with the two proposed multi-stage approaches.
    First, we observe that CEs fine-tuned with two-stages are effective BM25 re-rankers, but are on par with ColBERTv2 when it comes to re-rank its candidates.
    
    Next, we observe that the differences between D$\to$C and $C\to$D are not statistically significant for both CEs. However, to answer RQ2 despite this, D$\to$C appears to perform better than C$\to$D for Electra, and C$\to$D better than D$\to$C for RoBERTa.

    Lastly, we explore RQ3: whether is it a more effective re-ranker, a cross-encoder fine-tuned with the best single-stage fine-tuning strategy, or the best multi-stage one.
    ~\cref{tab:all_comparison} compares the effectiveness of CEs fine-tuned with the best one-stage and multi-stage fine-tuning approaches.
    We observe that although some improvements in effectiveness may seem considerable, there is no statistical difference between CEs fine-tuned with one stage or two.
    Also, across the different re-ranking benchmarks, multi-stage and single-stage fine-tuning yield to CEs with similar performances w.r.t. the baseline.
    To answer RQ3: there is no clear advantage in using two fine-tuning stages over one. Therefore, we conclude that a single stage of fine-tuning is sufficient for producing effective CE re-rankers.
    \begin{table}[t!]
    \centering
   \caption{Re-ranking effectiveness of CEs fine-tuned with best one-stage and multi-stage approaches (C for both, D$\to$C for El., C$\to$D for Ro.). Significant differences between the two fine-tuned versions of the same CE are denoted with \sd, and statistically significant differences w.r.t. the baseline are denoted with \dg.  Bold values denote the best value between two versions of the same CE, while \lw\ denotes values below the baseline. 
   } 
   \label{tab:all_comparison}
\resizebox{\textwidth}{!}{%
\begin{tabular}{@{}clcccccccccccc@{}}
\toprule
\multicolumn{2}{c}{\multirow{2}{*}{Re-rank}} & \multicolumn{3}{c}{DL 19}                                   & \multicolumn{3}{c}{DL 20}                           & \multicolumn{3}{c}{DL HARD}                            & \multicolumn{3}{c}{DEV SMALL}                     \\ 
\cmidrule(lr){3-5}\cmidrule(lr){6-8}\cmidrule(lr){9-11}\cmidrule(lr){12-14}
                     &                         & AP & nDCG        & MRR            & AP                & nDCG        & MRR         &   AP & nDCG        & MRR            & AP                & nDCG        & MRR        \\ 
\midrule
\multicolumn{14}{c}{BM25} \\
\midrule
\multicolumn{2}{c}{\bc --}                   & \bc.3035 & \bc.5121          & \bc.7138          & \bc.2811          & \bc.4769          & \bc.6653          & \bc.1622          & \bc.2886          & \bc.4740          & \bc.1941          & \bc.2301          & \bc.1855           \\
\multirow{2}{*}{El.} & C                    & \textbf{.3651}               & .7236\dg          & \textbf{.8314}\dg & .4012\dg          & .6759\dg          & .8278\dg          & \textbf{.2102}    & .3829\dg          & \textbf{.5197}    & .3689\dg          & .4203\dg          & .3709\dg           \\
                     & D$\to$C              & .3633                        & \textbf{.7304\dg} & .8262             & \textbf{.4065\dg} & \textbf{.6822\dg} & \textbf{.8438\dg} & .2050             & \textbf{.3841}\dg & .5041             & \textbf{.3693\dg} & \textbf{.4208\dg} & \textbf{.3712\dg}  \\ 
\midrule
\multirow{2}{*}{Ro.} & C                    & \textbf{.3687}               & \textbf{.7356}\dg & \textbf{.8651}    & .3997\dg          & .6720\dg          & \textbf{.8438}\dg & \textbf{.2230}    & \textbf{.4058}\dg & \textbf{.5657}    & \textbf{.3670}\dg & \textbf{.4182}\dg & \textbf{.3680}\dg  \\
                     & C$\to$D              & .3652                        & .7234\dg          & .8391             & \textbf{.4003}\dg & \textbf{.6775}\dg & .8380\dg          & .2228             & .4051\dg          & \textbf{.5657}    & .3667\dg          & \textbf{.4182}\dg & .3679\dg           \\ 
\midrule
\multicolumn{14}{c}{ColBERTv2}                                                                                                                                                                                                                                                                                              \\ 
\midrule
\multicolumn{2}{c}{\bc --}                   & \bc.5077 & \bc.7369          & \bc.8876          & \bc.5160          & \bc.7328          & \bc.8282          & \bc.2641          & \bc.4021          & \bc.5531          & \bc.3956          & \bc.4569          & \bc.3907           \\
\multirow{2}{*}{El.} & C                    & .4701\lw                     & .7537             & \textbf{.8663\lw} & .5205             & .7337             & .8536             & .2541\lw          & .4022             & .5150\lw          & .4228\dg          & .4844\dg          & .4191\dg           \\
                     & D$\to$C              & \textbf{.4732\lw}            & \textbf{.7632}    & .8599\lw          & \textbf{.5265}    & \textbf{.7585}    & \textbf{.8824}    & \textbf{.2552\lw} & \textbf{.4104}    & \textbf{.5254\lw} & \textbf{.4234\dg} & \textbf{.4855\dg} & \textbf{.4193\dg}  \\ 
\midrule
\multirow{2}{*}{Ro.} & C                    & \textbf{.4633\lw}            & .7333\lw          & .8391\lw          & .5136\lw          & .7370             & \textbf{.8617}    & \textbf{.2638\lw} & .4211             & .5640             & \textbf{.4151}\dg & \textbf{.4773}\dg & .4105\dg           \\
                     & C$\to$D              & \textbf{.4633\lw}            & \textbf{.7341\lw} & \textbf{.8411\lw} & \textbf{.5150\lw} & \textbf{.7375}    & \textbf{.8617}    & .2637\lw          & \textbf{.4217}    & \textbf{.5647}    & .4148\dg          & .4771\dg          & \textbf{.4106}\dg  \\
\bottomrule
\end{tabular} 
} 
    \end{table}

\section{Conclusions}

    In this work, we investigated the effectiveness of cross-encoders fine-tuned as point-wise re-rankers with single-stage and multi-stage approaches. Specifically, we compared models fine-tuned with a single stage of contrastive learning or distillation, and models further fine-tuned with the other approach.
    While fine-tuning with contrastive learning yields more effective re-rankers than with distillation, further refining fine-tuned models with a second stage yields no additional benefit. Our findings suggest that single-stage fine-tuning is sufficient for obtaining effective cross-encoder re-rankers. Future work could explore other contrastive learning and knowledge distillation losses, as well as  other training datasets, configurations, and families of neural re-rankers.

\begin{acks}
This work was partially supported by the Spoke ``FutureHPC \& BigData'' of the ICSC – Centro Nazionale di Ricerca in High-Performance Computing, Big Data and Quantum Computing funded by the Italian Government, the FoReLab and CrossLab projects (Departments of Excellence), the NEREO PRIN project funded by the Italian Ministry of Education and Research and European Union - Next Generation EU (M4C1 CUP 2022AEF-HAZ), and the FUN project (SGA 2024FSTPC2PN30) funded by the OpenWebSearch.eu project (GA 101070014).
\end{acks}

\bibliographystyle{splncs04}
\bibliography{bibliography}

\end{document}